# Accelerating Structures

*A. Degiovanni*
CERN, Geneva, Switzerland

**Abstract**
In this lecture the basic concepts of electromagnetic waves in accelerating structures are discussed. After a short introduction on the propagation of electromagnetic waves and the concept of travelling-wave and standing-wave structures, the most important parameters and figures of merit used to describe the acceleration efficiency and the performance of accelerating structures in terms of power consumption are presented. The process of radio-frequency design optimization is also discussed. Finally, a review of several types of structure using different accelerating modes is given, with examples of interest for medical accelerators.

## 1 Introduction

The aim of this lecture on 'accelerating structures' is to introduce the basic concepts and ideas related to the design and use of a radio-frequency (RF) accelerating system in linacs to an audience with backgrounds in several different fields. It does not cover the discussion of RF systems used for other types of machines (such as cyclotrons or synchrotrons, which are discussed in other lectures of these proceedings) and it does not pretend to be a detailed discussion of an extremely vast and complex field of accelerator physics and engineering. For the interested reader, a detailed introduction and description of this subject can be found in the Bibliography.

Accelerating structures are metallic resonant cavities used to accelerate beams of charged particles (i.e. to increase their energy). The acceleration of a charged beam is obtained by the interaction of the particles with the electromagnetic (EM) field confined in such void spaces delimited by metallic boundaries.

The force experienced by a particle of charge $q$ passing through an electromagnetic field with a certain velocity $v$ is described by the following equation:

$$\frac{d\vec{p}}{dt} = q\left(\vec{E} + \vec{v} \times \vec{B}\right). \tag{1}$$

The particle velocity $\mathbf{v} = v_x \mathbf{i} + v_y \mathbf{j} + v_z \mathbf{k}$ can be approximated—in the paraxial approximation (i.e. when $v_z \gg v_x, v_y$)—by $\mathbf{v} = \beta c\, \mathbf{k}$. In this case, it is clear that only the electric field can change the particle momentum along the $z$-axis. So in order to accelerate a charged beam along the structure axis, a longitudinal component of the electric field $E_z$ is needed.

## 2 Electromagnetic waves in RF structures

Electromagnetic waves are described by Maxwell's equations. In free space, they propagate as spherical waves and the intensity of the electromagnetic field decays as $1/r$, where $r$ is the distance from the point-like source to the measurement point. By using cylindrical and rectangular pipes, called waveguides, electromagnetic waves can propagate with very small losses.

In a wave-guiding system, the electric and magnetic fields are solutions of Maxwell's equations propagating along the guiding direction (the $z$ direction) and confined in the near vicinity of the guiding structure. They are mathematically described by the following equations:

$$E(x, y, z, t) = E(x, y)e^{j\omega t - jk_z z} , \qquad (2)$$

$$H(x, y, z, t) = H(x, y)e^{j\omega t - jk_z z} . \qquad (3)$$

They represent homogenous plane waves, characterized by the wave vector $\mathbf{k}$. The magnitude of the wave vector $\mathbf{k}$ is related to the EM-field angular frequency $\omega$ by the relation $k = \omega / c$. The wave vector $\mathbf{k}$ can be decomposed in its longitudinal and transverse components, respectively $k_z$ and $k_t$.

The precise relationship between $\omega$ and $k_z$ (generally called the *waveguide-propagation constant* or *wave number*) is called the *dispersion relation* and can be written as follows:

$$\omega = \sqrt{\omega_c^2 + k_z^2 c^2} . \qquad (4)$$

The dispersion diagram (or Brillouin diagram) shows the frequencies of electromagnetic waves that can be transmitted through a hollow conductor and is obtained by plotting Eq. (4) in a $\omega$–$k_z$ scatter plot. For rectangular and circular waveguides the dispersion relation has a hyperbolic shape as shown in Fig. 1.

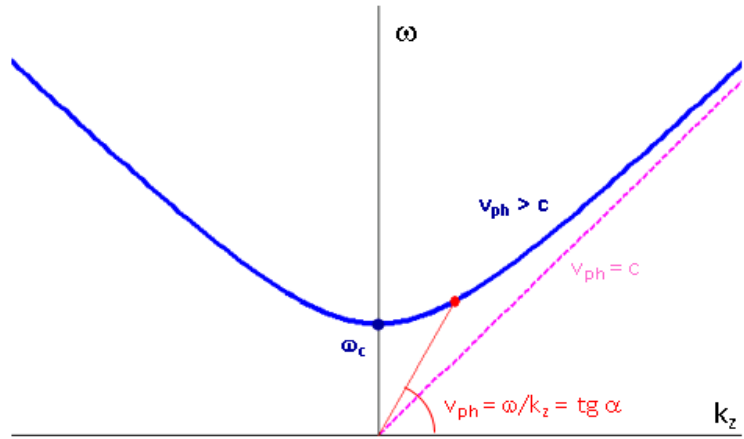

**Fig. 1:** Dispersion relation of a waveguide system

The term $\omega_c$ is the so-called *cut-off frequency*. Below the cut-off frequency, there is no propagation of the EM field. The boundary conditions for each waveguide type (i.e. the geometry of the inner section of the waveguide) force $\omega_c$ to take on only certain values. At each excitation frequency is an associated phase velocity, the velocity at which a certain phase travels in the waveguide.

The *phase velocity* is defined as the ratio between the excitation frequency $\omega$ (which is the angular frequency of the EM field once excited) and the waveguide-propagation constant $k_z$. This corresponds to the tangent of the angle between the line connecting the origin with the working point and the horizontal axis on a dispersion diagram:

$$v_\mathrm{ph} = \omega / k_z = \tan \alpha . \qquad (5)$$

In this condition, to be synchronized all the time with an accelerating $E$-field, a particle travelling inside the waveguide has to travel at $v = v_\mathrm{ph} > c$. This is clearly not possible and it will be shown in Section 2.2 how the phase velocity of the EM field can be reduced.

Energy and information travel at the group velocity $v_g = d\omega/dk$, which corresponds to the slope of the dispersion curve in the Brillouin diagram ($v_g < c < v_{ph}$).

## 2.1 Electromagnetic waves propagating modes

Solutions of Maxwell's equations can be classified in three families:

- TEM (Transverse Electric and Magnetic field) modes, where both electric and magnetic longitudinal components of the field are zero;
- TE (Transverse Electric) modes, where only the electric longitudinal component of the field is zero;
- TM (Transverse Magnetic) modes, where only the magnetic longitudinal component of the field is zero.

In bounded media (like a waveguide), the TEM modes are not allowed, because one of the field components must be in the direction of propagation to satisfy the boundary conditions. Only TM and TE modes are considered for propagation in bounded media. Subscripts are added to indicate different modes: $TM_{mnp}$ and $TE_{mnp}$. The meaning of these subscripts is different for rectangular and circular cavities:

- in rectangular cavities, the subscripts $m$ and $n$ are the number of half waves in the $x$ and $y$ directions, respectively. The additional subscript $p$ indicates the number of longitudinal half waves (i.e. in the $z$ direction);
- in circular cavities, the subscript $m$ indicates the number of full period variations of the field component in the azimuthal direction and $n$ is the number of zeros of the axial field component in the radial direction. The subscript $p$ indicates always the number of longitudinal half waves in the $z$ direction.

As already noted, in order to accelerate particles, a mode with a *longitudinal E-field* component on an axis is needed. A TM mode can be used for such purpose. Therefore, the simplest propagating mode is the $TM_{01}$.

## 2.2 Disc-loaded accelerating structure

A way to reduce the phase velocity of the $TM_{01}$ mode is to periodically add discs along the cylindrical waveguide, obtaining the so-called *disc-loaded waveguide*. The addition of discs inside the cylindrical waveguide, spaced by a distance $l$, induces *multiple reflections* between the discs. This action causes the dispersion curve to change (as shown in Fig. 2).

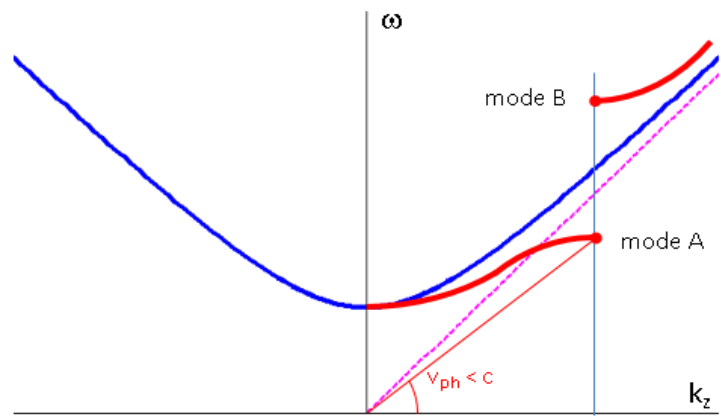

**Fig. 2:** Dispersion relation of a disc-loaded waveguide with stop band

In particular, for $k_z = 0$ or $k_z = \infty$ (corresponding to wavelength $\lambda_{ph} = \infty$ and $\lambda_{ph} = 0$), the wave does not see the effect of the discs. On the other hand, for waveguide wavelength $\lambda_{ph} = 2l$, the wave is confined between two discs and presents two polarizations (mode A and mode B in Fig. 2). These two modes have the same wavelength but different frequencies; the dispersion curve splits into two branches, separated by a *stop band*, where no propagation is allowed. Mode A can be used to accelerate particles since its phase velocity is smaller than the speed of light. On the other hand, mode B can propagate through the waveguide, but cannot be used to accelerate particles.

## 2.3 Travelling-wave and standing-wave structures

The disc-loaded waveguide is an example of a multi-cell structure. In such structures, a cell is defined by the space between two consecutive irises. The disc-loaded waveguide can be operated in two different conditions, i.e. as a travelling wave (TW) or a standing wave (SW).

In TW mode, the structure (or accelerating tube) has an input and output aperture (couplers) from which RF power can be, respectively, fed into the structure and extracted from it. The field propagates through each cell and the distance between the discs $l$ determines the phase advance between each cell:

$$\Delta\varphi = \frac{l}{\beta\lambda} 2\pi. \qquad (6)$$

When designing a linac, Eq. (6) is used to fix the length of the cells based on the choice of the frequency of excitation, the speed of the particle that needs to be accelerated, and the mode of operation.

SW modes are generated by the sum of two waves travelling in opposite directions, adding up in the different cells (Fig. 3). The boundary conditions at both ends impose that the electric field must be perpendicular to the reflecting plane. This results in the fact that only some modes on the disc-loaded dispersion curve are allowed. The resonant modes are characterized by the phase advance between each cell and depend on the number of coupled cells.

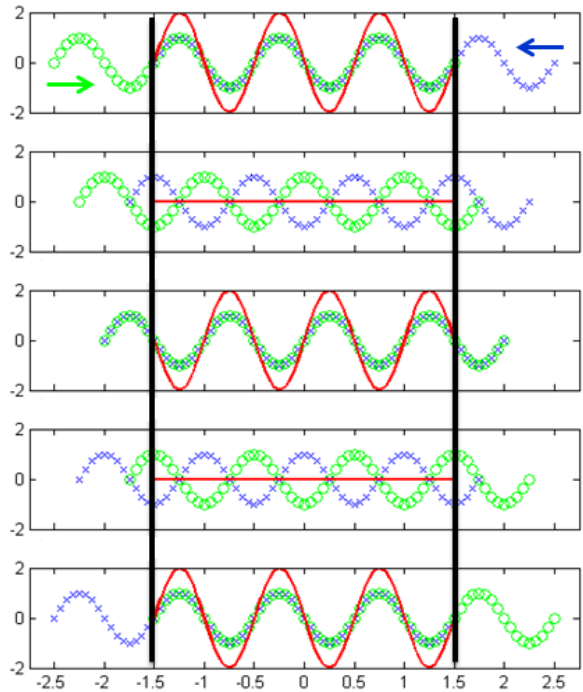

**Fig. 3:** Example of an SW pattern (red wave) obtained by superposition of two opposite travelling waves (blue and green curves). An SW cavity is obtained by inserting two metallic boundaries in correspondence of the two black lines.

Figure 4 shows the comparison of the dispersion curve of a multi-cell TW structure (left) and of a SW structure made of seven coupled resonators (right). As opposed to the case of TW structures, in SW structures there is no real power flow ($v_g = 0$) and, therefore, only one coupling port is needed for the excitation of the field inside the structure.

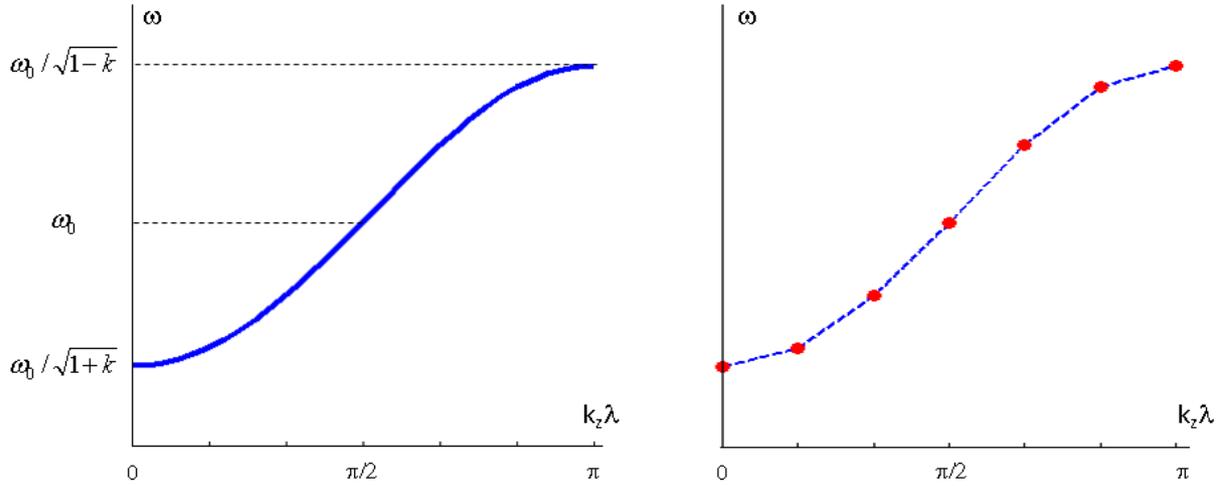

**Fig. 4:** Dispersion curve for coupled cavity oscillators with coupling factor $k$. In the case of TW mode (left), the dispersion curve is continuous (blue line), while for SW mode, only some modes are allowed (red dots).

The name of the resonant mode is typically given by the phase advance between consecutive cells. So, for example, a π-mode structure indicates a structure where the accelerating field is at maximum but in opposite directions in consecutive cells (it follows the scheme +1, −1, +1, −1, ...).

## 2.4 The pillbox cavity

A very important and instructive example of an accelerating structure is given by the so-called pillbox cavity, like the one shown in Fig. 5. It represents the simplest type of resonant electromagnetic structure. To use it as an accelerator cavity, one has to open two bore holes along the axis, introducing a small perturbation to the results obtained analytically [1].

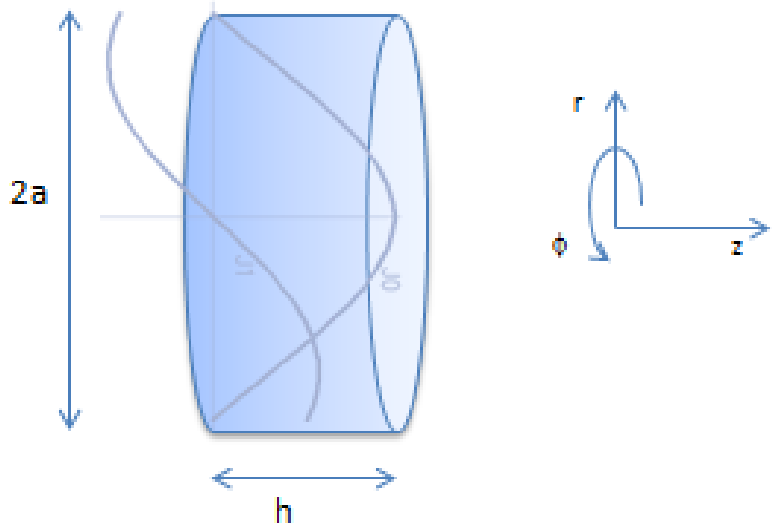

**Fig. 5:** Pillbox cavity, with longitudinal electric and azimuthal magnetic field components

For the simple geometry shown in Fig. 5, an analytical solution of the field components does exist and is given by the well-known Bessel's functions, $J$ and $J'$. In particular, the longitudinal component of the electric field and the azimuthal component of the magnetic field are described by

$$E_z = E_0 J_m(k_{mn} r) \cos(m\varphi) \cos\left(\frac{p\pi z}{h}\right) e^{j\omega t}, \tag{7}$$

$$B_\varphi = -j\omega \frac{a}{x_{mn} c^2} E_0 J'_m(k_{mn} r) \cos(m\varphi) \cos\left(\frac{p\pi z}{h}\right) e^{j\omega t}. \tag{8}$$

From a physical point of view, one can think that the boundary conditions on the cavity walls ($E_\parallel = 0$) force the fields to exist only at certain quantized resonant frequencies. The resonant condition is fulfilled when the wavelength is such that an integer number of half-wavelengths fits along each direction. The resonant frequency is given by the dispersion relation

$$\omega^2 / c^2 = k_{mn}^2 + k_z^2 = (x_{mn}/a)^2 + (p\pi/h)^2. \tag{9}$$

For the TM$_{010}$ mode, one has $m = 0$, $n = 1$, and $p = 0$, so (7) and (8) can be written as

$$E_z = E_0 J_0\left(\frac{2.405}{a} r\right) \cos(\omega t), \tag{10}$$

$$B_\varphi = -\frac{E_0}{c} J'_0\left(\frac{2.405}{a} r\right) \sin(\omega t). \tag{11}$$

## 3  Structure characteristics

Some of the fundamental properties of accelerating structures, which provide information about their performance, are described in the following. During the RF design phase, they are typically calculated using simulation tools that are able to solve Maxwell's equations inside complex geometrical structures. Some examples of such structures will be shown in Section 4. Depending on the application foreseen, one or more of such quantities are optimized by changing the geometry of the inner shape or by adding or removing features (for instance, by adding 'nose cones' close to the beam axis). It is important to underline that during the process of designing an accelerator, the RF optimization goes together with other considerations which can impose several constraints, such as beam dynamics, vacuum, beam instrumentations, and mechanical integration.

### 3.1  Energy gain in an accelerating gap

One of the most important quantities to evaluate for an accelerating cell of length $L$ is the *axial accelerating voltage*, defined as

$$V_{acc} = \int_0^L E_z e^{j\frac{\omega}{\beta c} z} dz. \tag{12}$$

The exponential factor accounts for the variation of the field, while particles with velocity $\beta c$ traverse the accelerating gap. The integral is taken over a distance $L$ in which the electric field is confined. It is very useful to express the accelerating voltage in the form

$$V_{acc} = V_0 T \tag{13}$$

where $V_0$ is the axial RF voltage or the voltage gain that a particle passing through a constant dc field equal to the field in the accelerating gap would experience, and $T$ is the so-called transit-time factor.

The *transit time factor* is the ratio of the acceleration voltage to the (non-physical) voltage a particle with infinite velocity would see. It is defined as follows:

$$T = \frac{|V_{acc}|}{\left|\int E_z dz\right|} = \frac{\left|\int E_z e^{j\frac{\omega}{\beta c}z} dz\right|}{\left|\int E_z dz\right|}. \tag{14}$$

The transit time factor describes the reduction in the energy gain caused by the sinusoidal time variation of the field in the accelerating gap. Its value ranges between 0 and 1.

The *energy gain* of an arbitrary particle with charge $q$ travelling through the accelerating gap is then given by

$$\Delta W = qV_0 T \cos\varphi. \tag{15}$$

Equation (15) is sometimes called Panofsky equation.

## 3.2  Acceleration efficiency figures of merit

Some of the power injected into the structure is lost or dissipated because of the electrical resistance in the cavity walls. The so-called *power losses*, or $P_{\text{loss}}$, are proportional to the *stored energy W*. In steady state, the total stored energy is

$$W = \iiint_{V_{\text{cavity}}} \left(\frac{\varepsilon}{2}|\vec{E}|^2 + \frac{\mu}{2}|\vec{H}|^2\right) dV. \tag{16}$$

The energy in the cavity is stored in the electric and magnetic field. Since $E$ and $H$ are 90° out of phase, the stored energy continuously swaps from electric energy to magnetic energy. The (imaginary part of the) Poynting vector describes this energy flux.

The ohmic losses can be evaluated from the surface resistance $R_s$ and the current density $J_s$ as

$$\frac{dP}{dA} = \frac{1}{2} R_s |J_s|^2. \tag{17}$$

The *surface resistance* is inversely proportional to the conductivity of the material $\sigma$ and to the so-called skin depth $\delta$, which represents the depth after which the field is attenuated by a factor 1/e:

$$R_s = \frac{1}{\sigma\delta}. \tag{18}$$

The value of the skin depth depends on the conductor material properties and scales as the square root of the inverse of the frequency:

$$\delta = \sqrt{\frac{2}{\omega\mu\sigma}}. \tag{19}$$

The cavity *quality factor Q* is defined as the ratio

$$Q = \frac{\omega_0 W}{P_{\text{loss}}} = \frac{\omega_0}{\Delta\omega}. \tag{20}$$

It describes the number of cycles needed to fill up or empty a structure at the resonant frequency of $\omega_0$. The ratio $P_{loss} / W$ can also be identified as the full width half maximum of the resonance peak $\Delta\omega$. High-$Q$ structures show, therefore, narrow band resonances and are thus typically very sensitive to frequency drifts due to temperature changes. It is important to notice that $Q$ is proportional to the stored energy.

The relation between gap voltage and power is characterized by the so-called *shunt impedance*:

$$R = \frac{V_0^2}{P_{loss}}. \tag{21}$$

By defining the average axial electric field over the cavity length $L$ as $E_0 = V_0 / L$, it is possible to introduce the *effective shunt impedance per unit length*:

$$ZT^2 = \frac{RT^2}{L} = \frac{E_0^2 T^2}{P_{loss}/L}. \tag{22}$$

Physically, it measures how well the RF power is concentrated in the useful region for acceleration. It is very important to underline that $ZT^2$ is *independent of the field level and cavity length*; it depends only on the cavity mode (frequency) and geometry (shape).

Another important quantity used to describe the structure efficiency is the ratio $R/Q$:

$$\frac{R}{Q} = \frac{\frac{V_0^2}{P_{loss}}}{\frac{\omega_0 W}{P_{loss}}} = \frac{V_0^2}{\omega_0 W}. \tag{23}$$

This quantity represents the proportionality constant between the square of the acceleration voltage and the stored energy. It is independent of cavity losses (it only depends on the geometry).

## 3.3 Field limiting quantities

High-gradient operation of linear accelerating structures is limited by a series of phenomena related to electrical discharges which are typically referred to as vacuum arcs or breakdowns. Although no theory or simulation method can predict breakdown performance of accelerating structures or high-power RF components during their technical design, a certain amount of experimental studies of the phenomena have been carried on in the past decades, mostly related to the development of high-gradient linear collider projects. From this experience, a series of scaling laws for high-gradient limiting quantities in terms of accelerating gradient and RF frequency has been discovered.

A quantity introduced in the late 1950s, but still used as a guideline for the design of structures at frequencies lower than 1 GHz, is the so-called Kilpatrik field limit. The Kilpatrik's criterion [2] for the determination of a threshold surface electric field defining the border between 'no vacuum sparking' and 'possible vacuum sparking' is described by

$$f = 1.64 \cdot E_K^2 \cdot e^{-\frac{8.5}{E_K}}, \tag{24}$$

where $f$ [MHz] is the RF frequency and $E_K$ [MV/m] is the Kilpatrick limit for the surface electric field.

The value of the surface electric field on the structure wall is proportional to the average gradient $E_0$ in the structure and one can define the peak to average field ratio $E_{S,max}/E_0$ of a cavity as the ratio between the maximum surface electric field $E_{S,max}$ and the average axial electric field $E_0$. The typical value for electron linacs is 2, while for proton and hadron linacs, depending on the design of the cell, this ratio can take values up to 4–5.

For a long time, the surface electric field has been considered to be the only limiting quantity for high-gradient operations. More recently, experimental evidence supports the idea that a combination of electric and magnetic fields at the surface correlates well with the measured breakdown probability [3] and that the stress induced by the surface field on the crystalline structure of the conductive material could be used to explain the breakdown mechanisms [4].

### 3.4 Accelerating cavities optimization

When designing the RF structures of a linac, it is important to keep in mind the final goals and objectives, and the constraints.

For example, for the case of a linac for proton therapy, important design goals are the actual energy gain and the final dose rate. Such goals should be achieved by respecting some constraints. For example, the size of the machine should fit in an assigned space, or the power consumption should not exceed a certain amount. Other limitations can come from technical and very often economic considerations, such as, for example, a limited number of RF power sources to reduce the total cost, a minimum repetition rate (to obtain the desired dose rate) and a maximum repetition rate (due to thermal considerations and to limitation in the RF power systems), beam dynamics considerations of drift lengths between structures and space for focusing elements along the linac, etc. During the design optimization, the accelerator physicist, taking into account the goals and constraints, can propose design solutions that maximize some critical parameters of the machine.

An example is given by the optimization of the effective shunt impedance per unit length defined in (22). In fact, combining Eqs. (22) and (15), one can write the following relation:

$$\Delta W \propto \sqrt{ZT^2 \cdot P \cdot L}. \tag{25}$$

Equation (25) reveals that to achieve a certain energy gain one can reduce the power consumption of the linac $P$ and/or the length of the machine $L$ by increasing the effective shunt impedance. So, the maximization of the value of $ZT^2$ during the design phase results in a reduction of the cost of the machine. In order to increase the $ZT^2$, one can change the structure geometry.

For example, let us consider the inner shape of an accelerating structure with nose cones like the one sketched in Fig. 6. By changing the radius of the bore hole ($Rb$), one affects the concentration of the field on axis and therefore, the effective shunt impedance. Generally speaking, by reducing the bore hole aperture, one can increase the effective shunt impedance of the structure. This action, on the other hand, would result in the reduction of beam transmission due to the smaller aperture and therefore, would be in conflict with the constraints given by beam dynamics considerations.

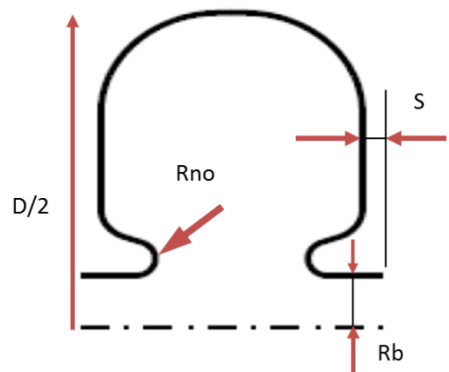

**Fig. 6:** Sketch of cell profile with parameters that can be used for the tuning and optimization

The results obtained during the optimization of the design for a structure resonating at 3 GHz, and designed for $\beta = 0.4$, are reported in Fig. 7. A reduction of 1 mm in the bore radius results in an increase of 10% of $ZT^2$, but the value of the bore radius should not go below a certain limit dictated by beam dynamics considerations. The same is true for a change of the septum thickness $S$. For a given cell length, by reducing the thickness of the walls between two cells, more space is left to the electric field and therefore, the $ZT^2$ increases. On the other hand, walls too thin would result in mechanical instabilities and problems in the cooling of the cells.

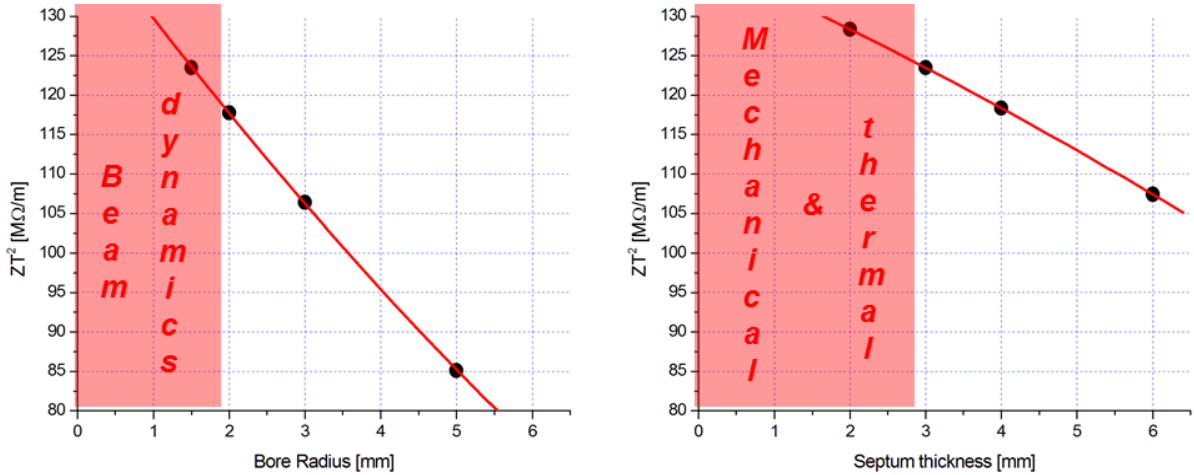

**Fig. 7:** Example of influence of the geometrical parameters on the effective shunt impedance

## 4 Example of structures

A large variety of structures are used in accelerator facilities worldwide. Typically, the structure design is adapted to the properties of the beam that is accelerated. A big difference exists between electron and proton or hadron linacs. In fact, due to the mass of 0.511 MeV/$c^2$, electron beams can be considered ultra-relativistic (β ≈ 1)—already at a kinetic energy of a few MeV. On the contrary, linacs are used for protons and hadrons up to relatively low energies of a few hundred MeV/u, corresponding to β ≈ 0–0.9. Since the accelerating efficiency is strongly dependent on the type of structure used and on the beam velocity, for proton and hadron linacs several structures are used in sequence to adapt to the increasing particle velocity, while electron linacs are typically made of the same type of structures.

Figure 8 shows the comparison of the values of effective shunt impedance calculated for eight different designs obtained in a joint study funded by EU called HIPPI (High Intensity Pulsed Power Injectors) [5]. The structures considered belong to two frequency ranges: 324–352 MHz, and its second harmonic 648–704 MHz. Similar results scaled towards larger shunt impedance are obtained with higher frequencies.

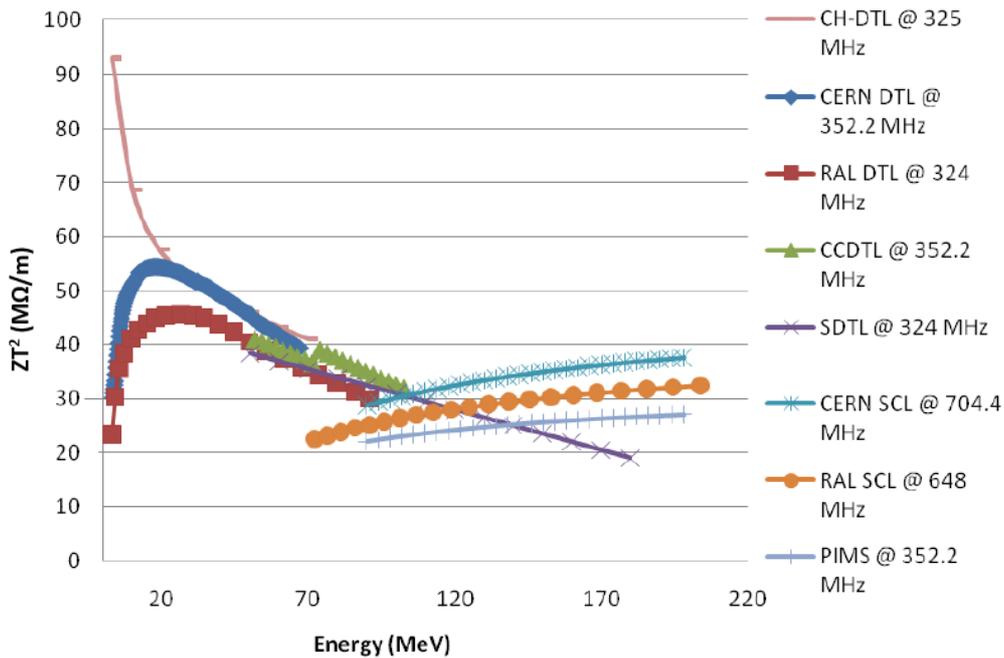

**Fig. 8:** Comparison of effective shunt impedance per unit length, for different types of structure, as a function of beam energy (taken from [5]).

Examples of structures using the TE mode are the Radio Frequency Quadrupole (RFQ) and the Interdigital and Crossbar H-mode (IH and CH) structures. Other structures using the TM mode are the Drift Tube Linac (DTL), the Cell Coupled Drift Tube Linac (CCDTL), the Cell Coupled Linac (CCL), and the elliptical cavities. In this section, several types of structures adapted to the acceleration of protons at different velocities are shown.

The curves of Fig. 8 present a strong dependence on beam energy. In the low-energy region, TE-mode-like structures are more efficient (CH structure), but above 20 MeV, their efficiency is comparable with that of TM-mode structures operating in 0-mode. Above 100 MeV, π-mode structures become more suitable.

### 4.1 TE-mode-like structures

The TE-mode-like structures (also called H-mode structures) are used at extremely low values of $\beta$. In this regime, focusing and bunching effects are more important than actual acceleration (see also the lecture on 'Beam dynamics' in these proceedings). The use of TE modes seems to be in contrast with what was discussed in Section 1, that only longitudinal electric field components are useful for acceleration. Indeed, pure TE modes have $E_z = 0$. In order to use them for acceleration, the field is forced to the longitudinal plane either by a longitudinal modulation (in the case of the RFQ) or by adding stems and drift tubes (in the IH or CH structures). Traditionally, H-mode structures have also been called Wideröe-linac structures.

#### 4.1.1 The Radio Frequency Quadrupole

The RFQ concept was first proposed by Kapchinskiy and Tepliakov in 1969 [6] and then further developed both in the Soviet Union and in the USA. In this type of structure, a quadrupolar electric field pattern is obtained by the presence of four poles which concentrate the electric field lines as shown in Fig. 9.

There are two types of RFQ construction geometries: the four vanes and the four rods. The typical frequencies used go from 10 to 400 MHz. Machining tolerances and related frequency errors limit the scaling up in frequency of such accelerators. The RFQ is placed just after the ion source with three main functions:

1. bunch the beam adiabatically, so as to prepare it for the next stage of acceleration with minimum losses;

2. focus the beam transversally by means of the electric quadrupolar field (extremely important at low energy when space charge forces are stronger);

3. accelerate the beam up to the few MeV required for injection into the next acceleration stage (typically a DTL).

Typical transmissions obtained with the RFQ are of 90–95%, but the real estate gradient is very low (typically no more than 1–2 MeV/m).

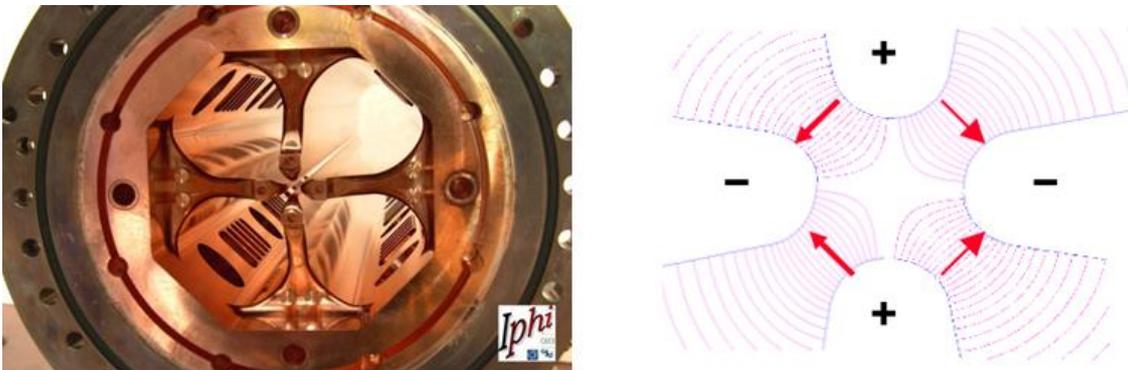

**Fig. 9:** An example of four vanes RFQ (left) with detail of the electric field lines close to the tips (right). The quadrupole field profile produces a focusing effect.

A novel design has been proposed at CERN for a high-frequency RFQ for medical application [7]. The RFQ is designed for an RF frequency of 750 MHz (almost double the maximum frequency used up to now), which makes it very compact compared to other similar machines (only 2 m for 5 MeV energy). It will be used as injector for proton-therapy linacs, but also its use as a booster for radio-isotope production is foreseen. A prototype is now being built at CERN which will be tested at the beginning of 2016.

*4.1.2    IH and CH structures*

After the first few MeV of acceleration, Interdigit and Crossbar H-mode structures (IH and CH) are more efficient. From the RF point of view, they are using, respectively, the $TE_{110}$ and $TE_{210}$ mode. However, the addition of the stems and the drift tubes forces the electric field (which in pure TE mode does not have a longitudinal component) along the beam axis, as is visible from the sketch in Fig. 10.

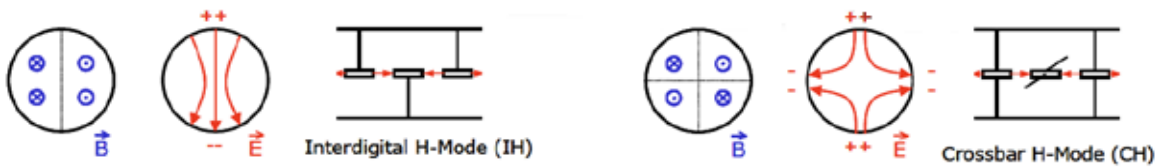

**Fig. 10:** Schematic representation of the mode excited in IH (left) and CH (right) structures (courtesy of F. Gerigk-CERN).

The H-mode structures have a good effective shunt impedance per unit length $ZT^2$ for $\beta$ in the range 0.02–0.08 and typically work at low frequencies (about 200 MHz). They are particularly well suited for low-intensity beams and they are used, for instance, in linac injectors for synchrotrons used in hadron therapy. For example, Fig. 11(left) shows the inner geometry of the IH structure used at CNAO (Pavia-Italy). The stems holding the drift tubes are placed alternately on two sides of the structure. The place reserved for Permanent Magnetic Quadrupole (PMQ) triplets is also visible. This machine is part of the injector chain and accelerates the particle from 400 keV/u to 7 MeV/u. In Fig. 11(right), the design of a CH structure, working at 3 GHz, to be used as proton booster from 15 to 66 MeV is shown. This structure, named CLUSTER (Coupled-cavity Linac USing Transverse Electric Radial field) [8], has a higher shunt impedance than other structures in this energy range.

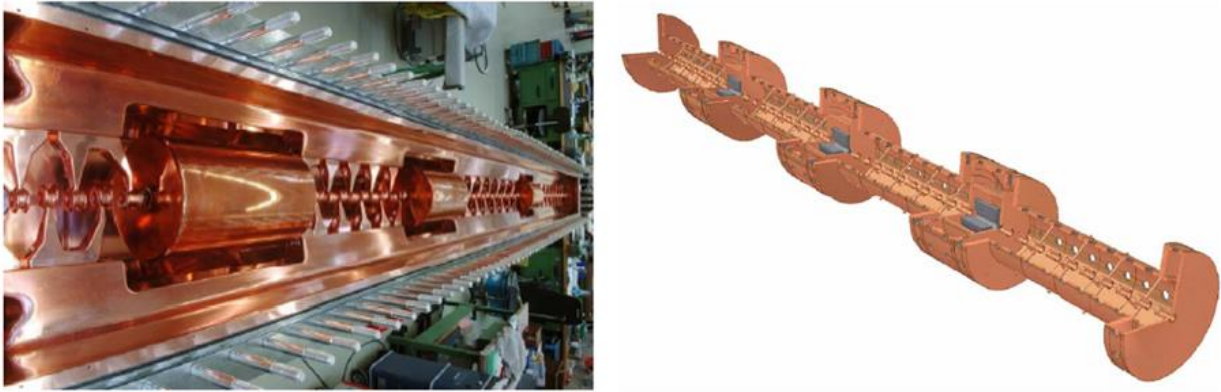

**Fig. 11:** Examples of IH (left) and CH (right) structures

In H-mode structures, the small drift tubes do not allow the insertion of quadrupoles. The focusing is provided by PMQ placed between tanks. A special beam dynamics approach is typically used (see also the lecture on 'Beam dynamics in linacs' in these proceedings).

### 4.2 TM-mode-like structures

The TM-mode structures, sometimes referred to as E-mode structures, are mostly used at intermediate $\beta$ values ranging from a few MeV up to several hundred MeV.

#### 4.2.1 The Drift Tube Linac

The DTL, also known as the Alvarez linac, is one of the simplest and oldest linac structures in use for the acceleration of protons and hadrons. It is a SW linac structure, used typically for $\beta$ in the range 0.1–0.4 and with RF frequencies of 20–400 MHz.

The name comes from the presence of drift tubes, held by stems (as shown in Fig. 12). In a certain sense, the inner geometry is similar to an IH structure, but the DTL is working in the $TM_{010}$ mode. The drift tubes are used to shield the particles from the RF field when this is in the decelerating phase. The length of the drift is such that when the particles are in the gaps between the drift tubes, they always see an accelerating field. It can be useful to consider the DTL as a multi-cell structure. The coupling between consecutive cells is at maximum, due to the fact that there are no walls with slots, so that the separation between cells is fully open. The 0 mode allows a long enough cell ($l = \beta\lambda$) to house focusing quadrupoles inside the drift tubes.

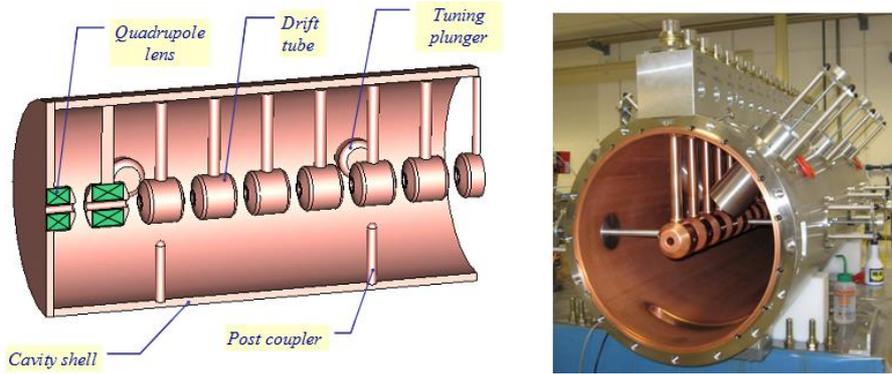

**Fig. 12:** Examples of the DTL module for LINAC4 at CERN (courtesy of M. Vretenar – CERN)

### 4.2.2 *The Cell Coupled Linac*

A CCL structure is made of a linear array of resonant cavities coupled together into a multi-cell structure.

In a normal CCL structure, the maximum frequency stability is obtained with the π/2 mode. This resonant mode is the least sensitive to frequency errors because it shows the largest neighbour mode separation. The synchronicity condition for acceleration is fulfilled for cavity length $l = \beta\lambda/4$. In this configuration (like the one shown in Fig. 13), each cavity has the same length, but only half of the cavities can be used for acceleration. The other half consist of unexcited cells. Consequently, the shunt impedance is lower than for a π-mode structure.

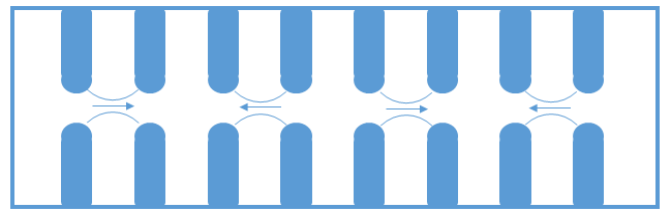

**Fig. 13:** Sketch of a CCL π/2-mode periodic structure

In order to keep the advantages of the π/2 mode in terms of frequency stability, and to achieve high shunt impedance, one can create a bi-periodic array of oscillators. For instance, one can reduce the space occupied axially by the unexcited cells—also called Coupling Cells (CCs)—and optimize the excited cells—or accelerating cells (ACs)—for higher shunt impedance. This is shown in Fig. 14 (left). This solution is typically called an *on-axis-coupled structure*.

A second type of geometry is the *side-coupled structure*, where the CCs are moved off-axis leaving all the axial space available for the ACs (Fig. 14 (right)). In this case the length of each AC is given by $l = \beta\lambda/2$.

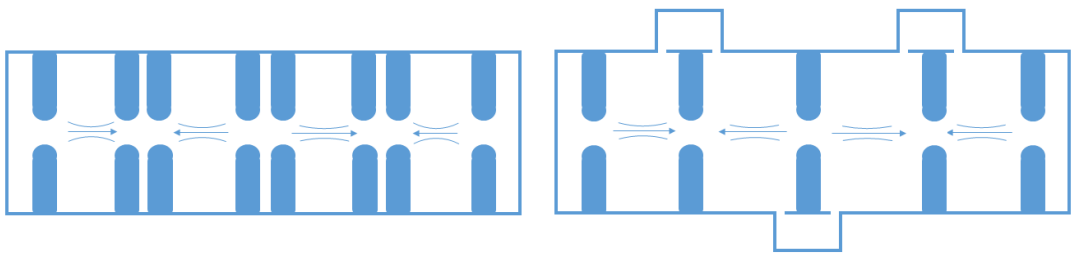

**Fig. 14:** Sketch of biperiodic CCL structures: on-axis-coupled structure (left) and side-coupled structure (right). The second case maximizes the space used for acceleration.

There are several types of CCL structures that are used to accelerate protons with $\beta$ between 0.4 and 0.9. Depending on the application each of them has advantages and disadvantages. An example of a 3 GHz Side Coupled Linac (SCL) structure designed to accelerate protons from 62 to 74 MeV is shown in Fig. 15. This module, called LIBO, was designed, constructed, and tested by TERA Foundation, in collaboration with CERN and INFN, and was the first 3 GHz linac to be used for the acceleration of a proton beam.

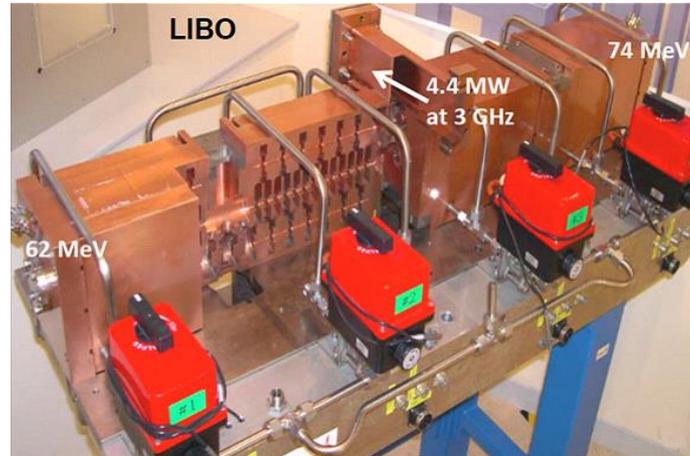

**Fig. 15:** The LIBO-module prototype is an example of 3 GHz SCL for proton therapy applications

### 4.2.3 *The Cell Coupled Drift Tube Linac*

A particular case of use of the CCL concept is given by the CCDTL. This is a kind of hybrid between a 0-mode DTL structure and a $\pi/2$-mode structure. In this case, short DTL tanks operating in 0 mode are coupled together via coupling cells, like in a CCL structure. The advantage of this structure, which is convenient for energies between 20 MeV and 100 MeV, is the longitudinal field stability given by the $\pi/2$ mode and the possibility of placing the focusing elements between the DTL tanks and not inside the drift tubes that, at low energies, are too short.

Two examples of CCDTL are shown in Fig. 16. Both are Side Coupled Drift Tube Linacs (SCDTL) due to the fact that the coupling cells are moved off-axis (on the side). On the left, a 352 MHz SCDTL module for the LINAC4 project at CERN, with two DTL tanks and one side coupling cell. On the right, the first module of a 3 GHz SCDTL designed by ENEA (Frascati-Italy) for a proton-therapy linac project.

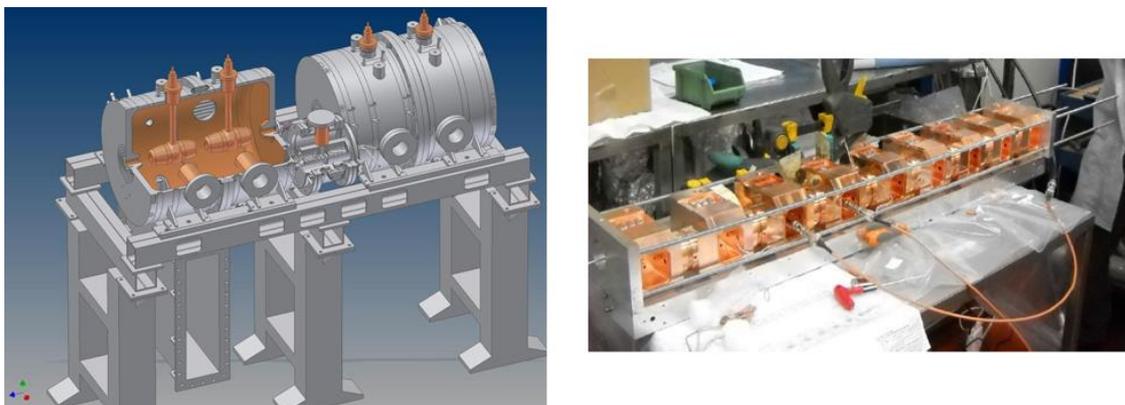

**Fig. 16:** Two examples of CCDTL structures: on the left is one of the 352 MHz SCDTL modules for LINAC4 (courtesy of M. Vretenar – CERN), and on the right is a 3 GHz SCDTL designed by ENEA for a proton therapy linac (courtesy of L. Picardi – ENEA).

## 4.3 Summary of linac structures

The following table (Table 1) summarizes the properties of the different linac structures discussed in this section for proton-therapy applications. This is given only as a general reference.

Table 1: Summary table of linac structures for protons

| Structure type | Family | Mode | Typical energy [MeV] |
| --- | --- | --- | --- |
| Radio Frequency Quadrupole | TE | $TE_{21}$+vanes | 0–2 |
| Interdigit H-mode (Wideroe) | TE | $TE_{110}$ | 0.4–10 |
| Crossbar H-mode (Wideroe) | TE | $TE_{210}$ | 0.4–10 |
| Drift Tube Linac (Alvarez) | TM | 0 mode | 4–80 |
| Cell Coupled Linac | TM | $\pi/2$ mode | 80–400 |
| Cell Coupled Drift Tube Linac | TM | Hybrid | 20–100 |

## 5 RF linacs for medical applications

More than 15,000 electron linacs are used daily for radiotherapy treatment. They represent about 50% of all the accelerators with energies greater than 1 MeV and their number is continuously growing [9]. Electron linacs for radiotherapy can be considered industrial products.

A linac radiotherapy unit includes not only the accelerator, but is composed of the RF power source, the supporting structure that usually can rotate around the patient (gantry), the couch, the alignment system, and the control system, typically integrated with the treatment planning system. For what constitutes the accelerator part, electron tubes can be either SW or TW. For example, some manufacturers propose SCL type structures and others propose disc loaded waveguides. As RF sources, magnetrons or klystrons in the 5 MW range are used. The typical frequency used is 2.856 GHz in the USA and 2.998 GHz in Europe.

Recently, a CERN spin-off company called A.D.A.M. (Application of Detector and Accelerators to Medicine) has started the design and production of a high-frequency (3 GHz) linac for proton therapy, with the idea to commercialize it. As is shown in Fig. 17, a modular approach has been chosen, with a sequence of three types of linac structures discussed in the previous section: an RFQ, followed by an SCDTL, and an SCL.

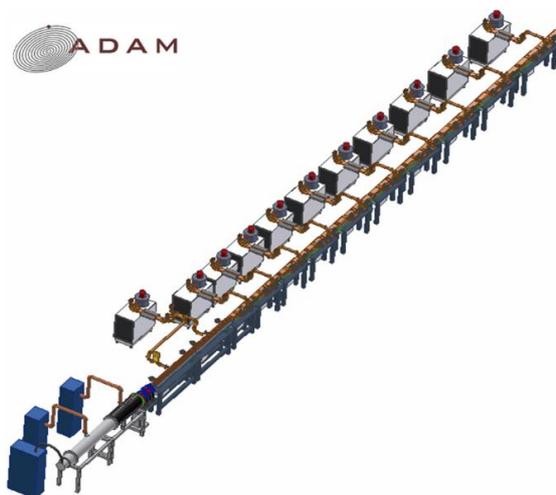

**Fig. 17:** Layout of a high-frequency linac for proton therapy, proposed by the CERN spin-off company A.D.A.M. (courtesy of A.D.A.M.).

**Acknowledgements**

Most of the material for this lecture comes from previous lectures of M. Vretenar, E. Jensen and F. Gerigk. My thanks to all of them.